\documentclass[twocolumn,showpacs,preprintnumbers,amsmath,amssymb,superscriptaddress]{revtex4-1}
\usepackage{graphicx}
\usepackage{dcolumn}
\usepackage{bm}

\begin{document}
\title{2D atom localization in a four-level tripod system in laser fields}
\author{Vladimir Ivanov}
\email{ivvl82@gmail.com} \affiliation{Turku Centre for Quantum
Physics, Department of Physics and Astronomy, University of Turku,
20014 Turku, Finland} \affiliation{Saint-Petersburg State
University of Information Technologies, Mechanics and Optics,
197101 St. Petersburg, Russia}
\author{Yuri Rozhdestvensky}
\email{rozd-yu@mail.ru} \affiliation{Saint-Petersburg State
University of Information Technologies, Mechanics and Optics,
197101 St. Petersburg, Russia}
\date{Received 5 September 2009; published 8 March 2010}

\begin{abstract}
We propose a scheme for two-dimensional (2D) atom localization in
a four-level tripod system under an influence of two orthogonal
standing-wave fields. Position information of the atom is retained
in the atomic internal states by an additional probe field either
of a standing or of a running wave. It is shown that the
localization factors depend crucially on the atom-field coupling
that results in such spatial structures of populations as spikes,
craters and waves. We demonstrate a high-precision localization
due to measurement of population in the upper state or in any
ground state.
\end{abstract}


\maketitle

\section{Introduction}
\label{s1}

During the last years, spatial localization of an atom using
optical techniques has attracted extensive attention. The
prototype of an all-optical measuring device is the Heisenberg
microscope \cite{H}. The resolving power of this particular
experiment is limited by the optical wavelength if the Heisenberg
uncertainty relation is taken into account. Further, in the
one-dimensional (1D) case, optical methods for atom localization
within the optical wavelength were developed using an internal
structure of the atomic levels.

Schemes of subwavelength localization have been performed, using
for example, the measurement either of the phase shift
\cite{Storey,Storey1,Storey2} or of the atomic dipole \cite{Kunze}
after the passage of the atom through an off-resonant
standing-wave field. Another method has been proposed by Thomas
and co-workers \cite{Thomas,Gardner,Gardner1} in which a spatially
varying potential correlates an atomic resonance frequency with
the atomic position. In addition, a few schemes for position
measurement were performed, using absorption light masks
\cite{Zeil,Zeil1,John} in which the precision of measurement
achieved, was nearly tens of nanometers.

Recently, subwavelength localization of an atom in a standing-wave
field was demonstrated in the case of resonant atom-field
coupling. For example, two localization schemes were proposed by
Zubairy and co-workers using either measurement of the spontaneous
spectrum in a three-level system or the resonant fluorescence in a
two-level system for localization of the atom during its motion in
the standing wave \cite{Zub,Zub1,Zub2,Zub3}. Another related
localization scheme based on measurement of the upper-state
population has been proposed by Paspalakis and Knight
\cite{PK,PK1} for a three-level $\Lambda$-type atom interacting
with two fields, a laser probe field and a classical standing-wave
field.

In this paper we demonstrate the two-dimensional (2D)
subwavelength localization of an atom in a four-level tripod
system due to interaction with a probe laser field and two
resonant standing-wave fields. In fact, most 2D level systems
consisting of a large number of levels do not match for a
subwavelength localization scheme because of interference among
states. On the other hand, tripod configuration of atomic levels
consists only of four levels involving three atomic ground states
and one excited state. We should point out that tripod schemes are
experimentally accessible in metastable Ne, $^{87}$Rb and a number
of other gases \cite{Th,Vew}.

The paper is organized as follows. In Section \ref{s2} we give the
equations of motion of the density-matrix elements in this
configuration which are valid for both the running-wave and
standing-wave cases of the probe field. In these equations, the
spontaneous decay from the upper state to ground states is taken
into account. Sections \ref{s3} and \ref{s4} demonstrate a
high-precision localization due to measurement of the upper-state
population as well as of a ground-state population. The spatial
structures of the populations take the form of spikes, craters or
waves depending on the atom-field coupling. Moreover, using
measurement of a ground-state population, the atom can be
localized at the nodes of the standing waves.

\section{Model and equations}
\label{s2}

The laser configuration consists of two parallel $\sigma_+$,
$\sigma_-$ polarized optical waves and an orthogonal linearly
polarized optical wave. In such a configuration the positive
frequency part of the electric field can be written as
\begin{equation}
\vec E^{(+)}(\vec r,t)=\vec E_1(x)e^{-i\omega_1 t}+\vec
E_3(x)e^{-i\omega_3 t}+\vec E_2(y)e^{-i\omega_2 t}.
\end{equation}
The first term corresponds to a $\sigma_+$ circularly polarized
wave with the frequency $\omega_1$ and the second one corresponds
to a $\sigma_-$ circularly polarized wave with the frequency
$\omega_3$. The third term describes a linearly polarized wave
with the frequency $\omega_2$. The electrical fields of the
standing waves are $\vec E_1(x)=\vec{\mathcal E}_1\sin kx$ and
$\vec E_2(y)=\vec{\mathcal E}_2\sin ky$, while the $\vec E_3(x)$
corresponds to an additional probe field. The probe field can be
either a running wave $\vec E_3(x)=\vec{\mathcal E}_3\exp(ikx)$
along the $x$ direction or a standing wave $\vec
E_3(x)=\vec{\mathcal E}_3\sin(kx+\psi)$ where $\psi$ is the
spatial phase shift between the standing waves $\vec E_1(x)$ and
$\vec E_3(x)$.

\begin{figure}
(a)\includegraphics[width=5cm]{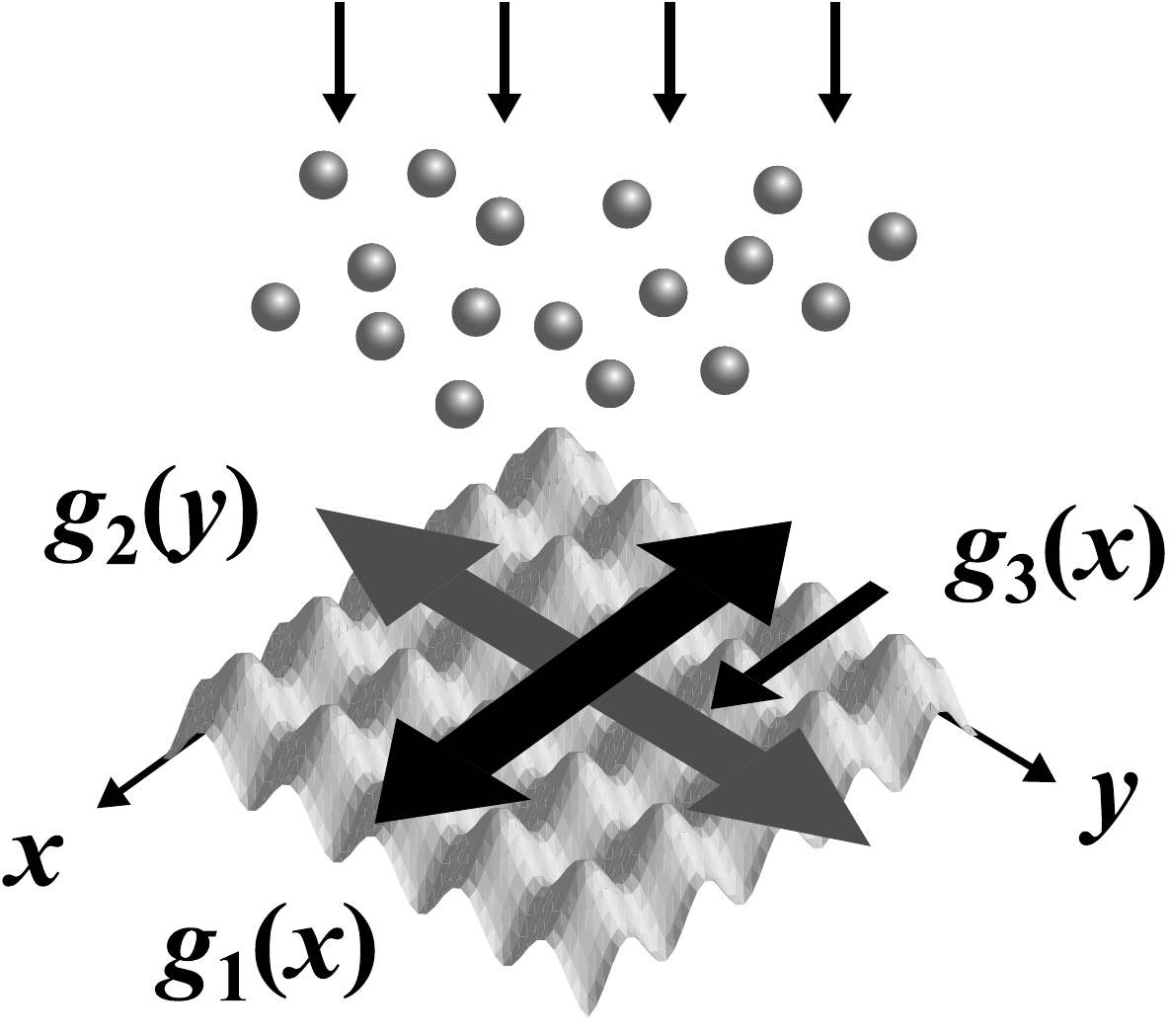}\\
(b)\includegraphics[width=6cm]{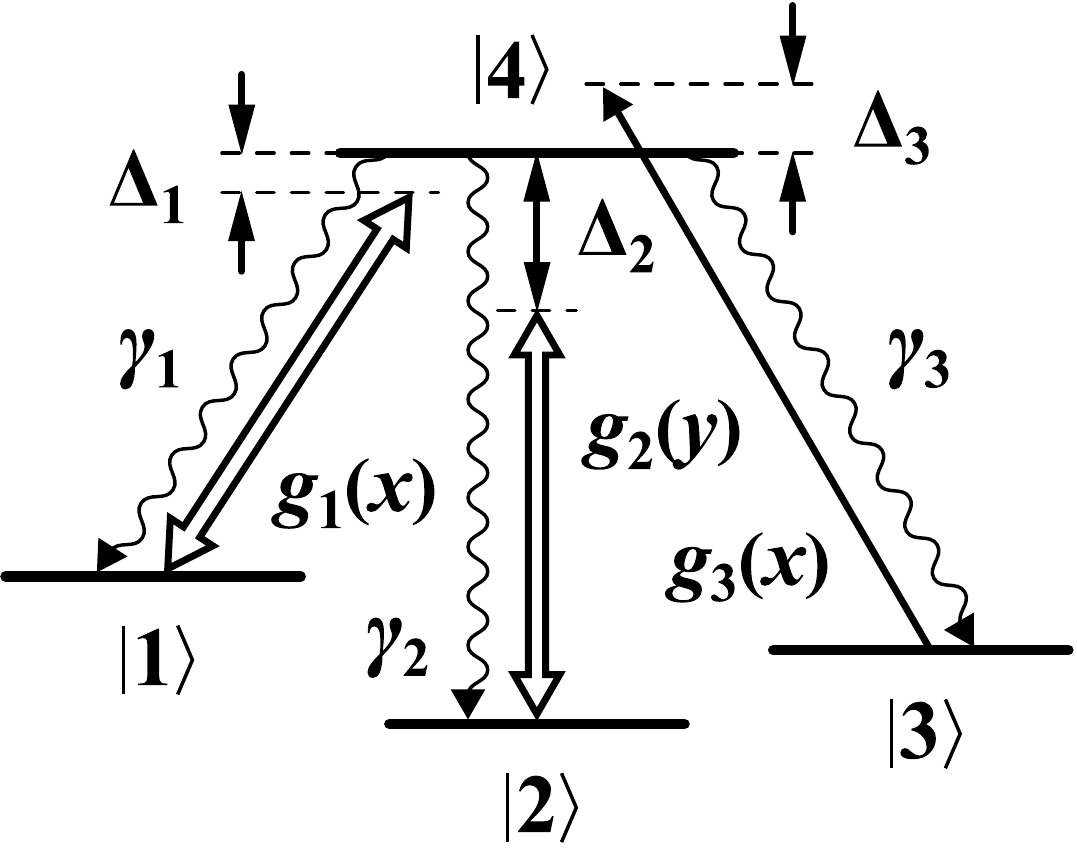}
\caption{\label{Tripod}(a) Atoms pass through an interaction range
of coupling with two standing-wave laser fields with the Rabi
frequencies $g_1(x)$, $g_2(y)$ and probe laser field with the Rabi
frequency $g_3(x)$. (b) A tripod scheme of atomic energy levels.
Standing-wave field $g_1(x)$ is nearly resonant with the
$|1\rangle\leftrightarrow |4\rangle$ transition and the second
standing-wave field---$g_2(x)$ is nearly resonant with the
$|2\rangle\leftrightarrow |4\rangle$ transition, while the probe
laser field interacts with the $|3\rangle\leftrightarrow
|4\rangle$ transition.}
\end{figure}

The atom moves parallel to the $z$ axis and passes through the
interaction range as seen in Fig. \ref{Tripod}(a). The tripod
scheme of the atomic levels which has the upper state $|4\rangle$
and the three ground states $|1\rangle$--$|3\rangle$ is shown in
Fig. \ref{Tripod}(b). The transition $|1\rangle\leftrightarrow
|4\rangle$ is taken to be nearly resonant with the standing wave
$\vec E_1(x)$ with the detuning $\Delta_1=\omega_1-\omega_{41}$,
while the transition $|2\rangle\leftrightarrow |4\rangle$ is taken
to be nearly resonant with the standing wave $\vec E_2(y)$ with
the detuning $\Delta_2=\omega_2-\omega_{42}$. The atom interacts
with the probe laser field $\vec E_3(x)$, which is resonant with
the $|3\rangle\leftrightarrow |4\rangle$ transition, with the
detuning $\Delta_3=\omega_3-\omega_{43}$.

The coupling of the atom with each of the strong standing waves is
characterized by the Rabi frequencies
\begin{equation}
g_1(x)=G_1\sin kx,\quad g_2(y)=G_2\sin ky,
\end{equation}
where $G_j=|\vec d_{4j}\vec{\mathcal E}_j/\hbar|$ ($j=1,2$). The
Rabi frequency of the probe field is
\begin{align}
g_3(x)=\left\{
\begin{array}{cl}
\Omega & \mbox{in a running-wave case},\\
\Omega\sin(kx+\psi) & \mbox{in a standing-wave case},
\end{array}
\right.
\end{align}
where $\Omega=|\vec d_{43}\vec{\mathcal E}_3/\hbar|$.

We assume that the center-of-mass position of the atom is nearly
constant along the directions of the laser waves and neglect the
kinetic part of the atom from the Hamiltonian in the Raman-Nath
approximation \cite{Meystre}. Note that during the interaction the
atomic position does not change directly and our localization
scheme only influences the internal states of the atom. Then, in
the interaction picture and the rotating wave approximation the
equations of matrix-density elements can be written as
\begin{subequations}\label{eq}
\begin{gather}
i\dot\rho_{11}=g_1(x)(\rho_{14}-\rho_{41})+i\gamma_1\rho_{44},\\
i\dot\rho_{22}=g_2(y)(\rho_{24}-\rho_{42})+i\gamma_2\rho_{44},\\
i\dot\rho_{33}=g_3(x)(\rho_{34}-\rho_{43})+i\gamma_3\rho_{44},\label{eq_c}\\
i\dot\rho_{44}=g_1(x)(\rho_{41}-\rho_{14})+g_2(y)(\rho_{42}-\rho_{24})\notag\\
+g_3(x)(\rho_{43}-\rho_{34})-2i\gamma\rho_{44},\\
i\dot\rho_{12}=g_2(y)\rho_{14}-g_1(x)\rho_{42}
+(\Delta_{12}-i\Gamma_{12})\rho_{12},\\
i\dot\rho_{13}=g_3(x)\rho_{14}-g_1(x)\rho_{43}
+(\Delta_{13}-i\Gamma_{13})\rho_{13},\label{eq_f}\\
i\dot\rho_{23}=g_3(x)\rho_{24}-g_2(y)\rho_{43}
+(\Delta_{23}-i\Gamma_{23})\rho_{23},\label{eq_g}\\
i\dot\rho_{14}=g_1(x)(\rho_{11}-\rho_{44})+g_2(y)\rho_{12}\notag\\
+g_3(x)\rho_{13}+(\Delta_1-i\Gamma_{14})\rho_{14},\\
i\dot\rho_{24}=g_2(y)(\rho_{22}-\rho_{44})+g_1(x)\rho_{21}\notag\\
+g_3(x)\rho_{23}+(\Delta_2-i\Gamma_{24})\rho_{24},\\
i\dot\rho_{34}=g_3(x)(\rho_{33}-\rho_{44})+g_1(x)\rho_{31}\notag\\
+g_2(y)\rho_{32}+(\Delta_3-i\Gamma_{34})\rho_{34}\label{eq_j},
\end{gather}
\end{subequations}
together with $\rho_{mn}=\rho_{nm}^*$,
$\sum\limits_{n=1}^4\rho_{nn}=1$ and
$\Delta_{mn}=\Delta_m-\Delta_n$. The atomic decay takes place from
the level $|4\rangle$ to the levels
$|1\rangle,|2\rangle,|3\rangle$ with the rates $\gamma_1$,
$\gamma_2$ and $\gamma_3$, respectively. The decay rates of atomic
coherences between the upper level and ground levels equal
$\Gamma_{14}$, $\Gamma_{24}$ and $\Gamma_{34}$. Coherent decay
rates $\Gamma_{12}$, $\Gamma_{13}$ and $\Gamma_{23}$ are usually
much less than the rates of decay through the
$|4\rangle\leftrightarrow |m\rangle$ ($m=1,2,3$) channels, so the
approximate equalities are
$\Gamma_{12}=\Gamma_{13}=\Gamma_{23}=0$.

\section{2D atom localization via the upper-level population}
\label{s3}

The dynamics of interaction vary depending on the position of the
atom which passes through laser fields in the $(x,y)$ position. In
case of a long-time interaction the final populations in internal
states are defined only by the laser fields and the atom does not
need an initial-state preparation. The final distribution of the
upper-level population depends on three controllable detunings
$\Delta_1$, $\Delta_2$ and $\Delta_3$, and on the Rabi frequencies
of the driving fields, $G_1$, $G_2$ and $\Omega$. Also, in a
standing-wave case of the probe field the distribution depends on
spatial shift $\psi$. Consequently, controlling the parameters of
this interaction allows us to create different patterns of the
upper-level population.

When the probe laser field is weak the measurement of the
upper-state population leads to a high-precision localization, as
shown in the related 1D scheme for a three-level $\Lambda$-type
atom \cite{PK,PK1}. We assume that $\Omega\ll
G_1,G_2,\gamma_m,\Gamma_{m4},|\Delta_m|,|\Delta_{mn}|$
($m,n=1,2,3)$ is satisfied and obtain the long-time population in
state $|4\rangle$ in second order of perturbation theory. Note
that the second-order perturbation theory is correct only when
$\Omega\ll g_1(x),g_2(y)$, which corresponds to the atomic
positions far from the nodes of standing waves $g_1(x)$ and
$g_2(y)$.

In the limit case of $\Omega=0$ the atomic population is
concentrated in the $|3\rangle$ state. Therefore, when the probe
laser field is weak, we have $\rho _{33}\approx 1$ during the
atom-field interaction. In long-time limit, $\dot\rho_{mn}=0$ and
the non-diagonal elements $\rho_{31}$ and $\rho_{32}$ can be
derived from (\ref{eq_f}) and (\ref{eq_g}) equations and are given
by
\begin{equation}\label{coh}
\rho_{31}\approx\frac{g_1(x)}{\Delta_{13}}\rho_{34},\quad
\rho_{32}\approx\frac{g_2(y)}{\Delta_{23}}\rho_{34}.
\end{equation}
Here we used a correlation $\rho^*_{ij}=\rho_{ji}$. Substitution
of Eq. (\ref{coh}) and an approximate expression
$\rho_{33}-\rho_{44}\approx 1$ into (\ref{eq_j}) gives the
non-diagonal element
\begin{equation}
\rho_{34}\approx\frac{g_3(x)}{i\Gamma_{34}-\dfrac{g_1^2(x)}{\Delta_{13}}
-\dfrac{g_2^2(y)}{\Delta_{23}}-\Delta_3}.
\end{equation}
Then, the upper-level population is obtained from Eq. (\ref{eq_c})
and is given by
\begin{equation}
\rho_{44}\approx -\frac{2g_3^2(x)}{\gamma_3}\mathop{\rm Im}
\frac{1}{i\Gamma_{34}-\dfrac{g_1^2(x)}{\Delta_{13}}
-\dfrac{g_2^2(y)}{\Delta_{23}}-\Delta_3}.
\end{equation}
This expression can be written as
\begin{equation}\label{final}
\rho_{44}\approx\frac{2\Gamma_{34}g_3^2(x)}{\gamma_3 Z},
\end{equation}
where
\begin{equation}\label{denom}
Z=\left(\frac{G_1^2}{\Delta_{13}}\sin^2
kx+\frac{G_2^2}{\Delta_{23}}\sin^2 ky
+\Delta_3\right)^2+\Gamma_{34}^2.
\end{equation}

\begin{figure}
\includegraphics[width=7.5cm]{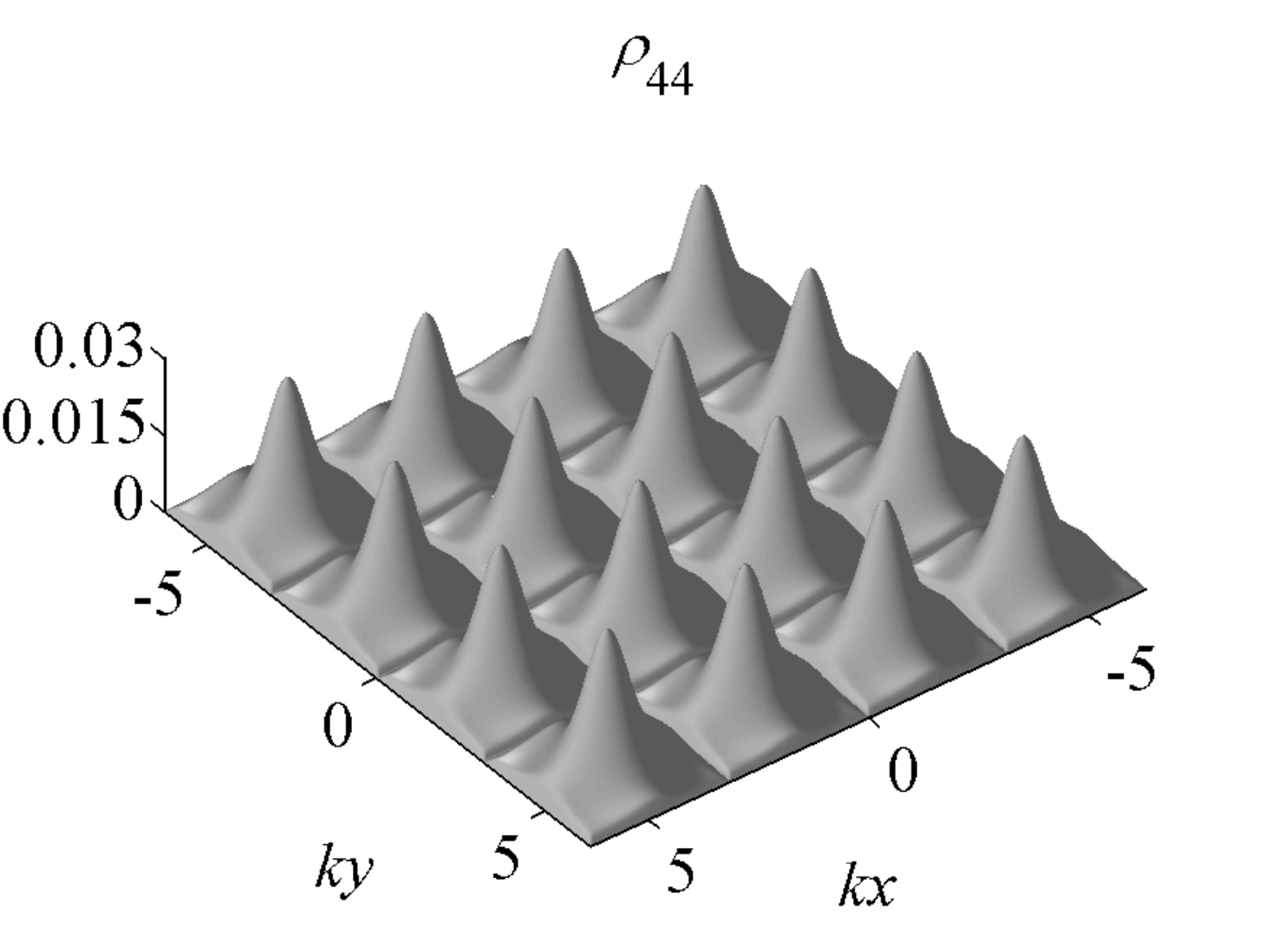}\\(a)\\
\includegraphics[width=7.5cm]{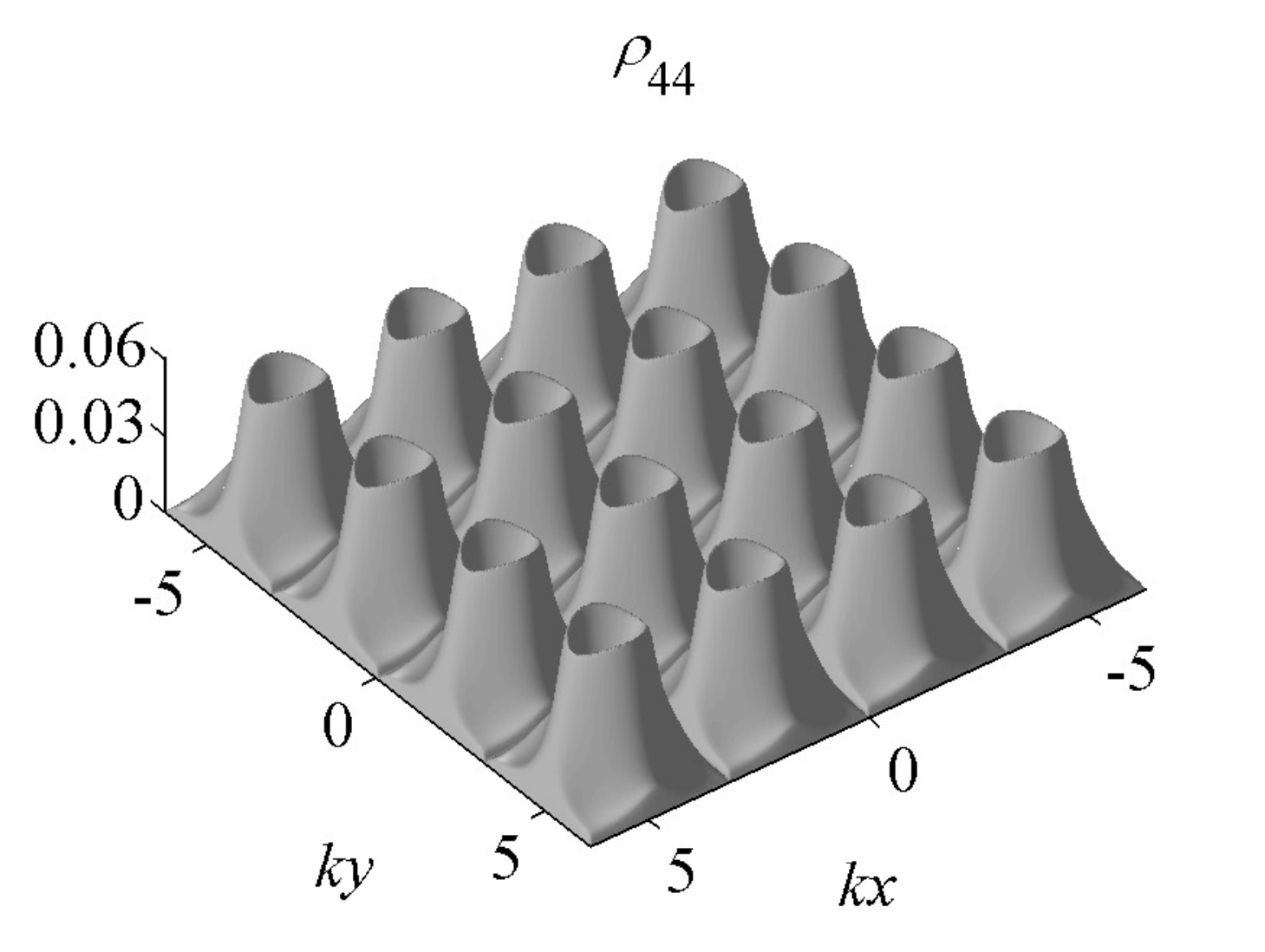}\\(b)\\
\includegraphics[width=7.5cm]{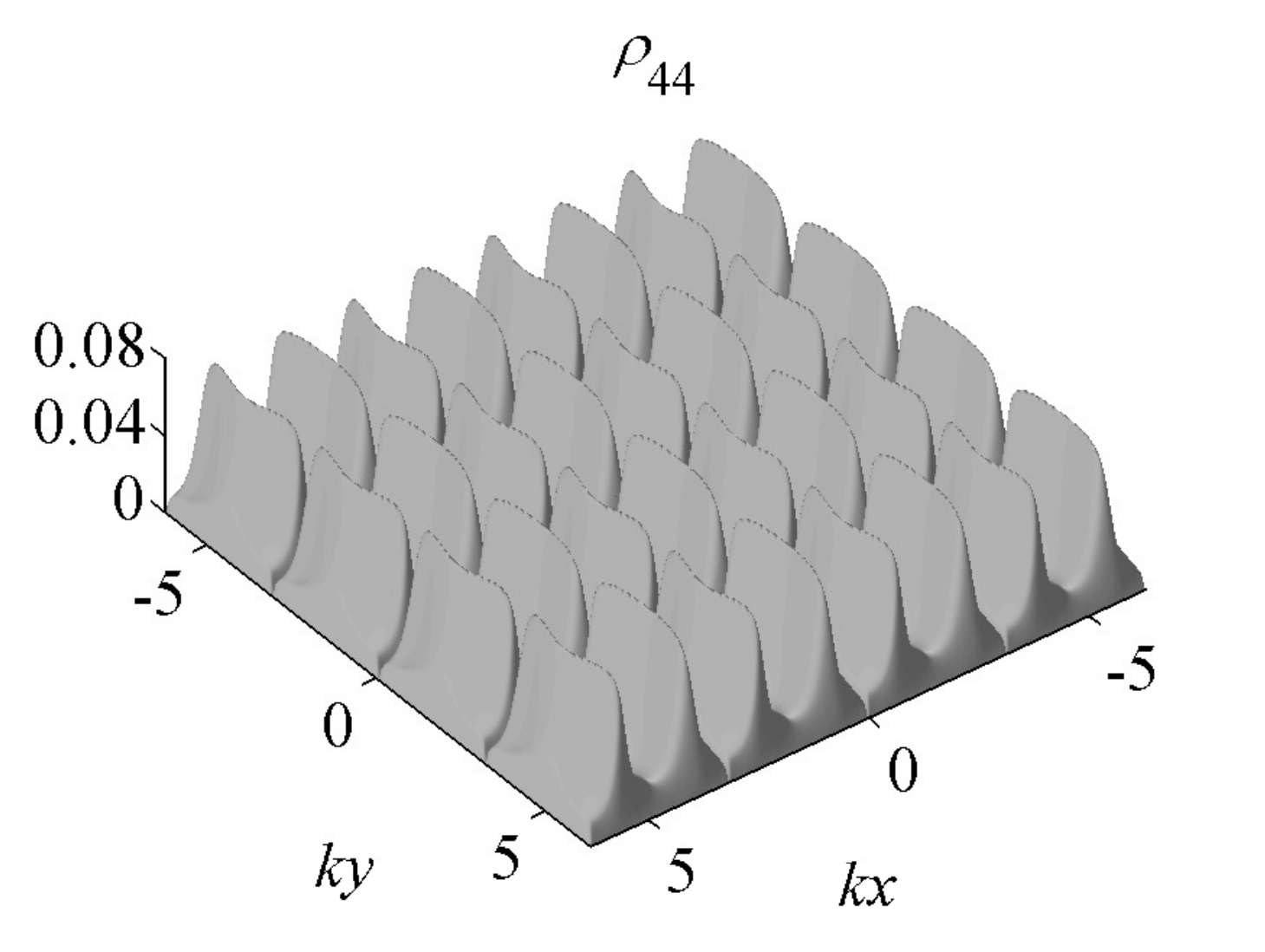}\\(c) \caption{\label{upper}
The upper-level population $\rho_{44}$ as a function of $(kx,ky)$.
The spatial distributions of the population represent such 2D
periodic structures as (a) spikes, (b) craters and (c) waves in
dependence on the laser frequency detunings, (a)
$\Delta_1=10\gamma$, $\Delta_2=14\gamma$, $\Delta_3=16\gamma$; (b)
$\Delta_1=5\gamma$, $\Delta_2=9\gamma$, $\Delta_3=11\gamma$; and
(c) $\Delta_1=\gamma$, $\Delta_2=-3\gamma$, $\Delta_3=5\gamma$.
The Rabi frequencies of the standing waves are $G_1=6\gamma$,
$G_2=4\gamma$, and the Rabi frequency of the probe running wave is
$\Omega=0.3\gamma$. Other parameters are as follows. Decay rates
$\gamma_{1,2,3}$ from the upper level equal $\gamma$,
coherent-decay rates through the channels
$|4\rangle\leftrightarrow |m\rangle$ ($m=1,2,3$) are
$\gamma_{14}=1.5\gamma$, $\gamma_{24}=1.5\gamma$,
$\gamma_{24}=1.5\gamma$, and coherent-decay rates among ground
states are $\gamma_{12}=0$, $\gamma_{13}=0$, $\gamma_{23}=0$.}
\end{figure}

The expressions (\ref{final}), (\ref{denom}) allows us to analyze
distribution of the upper-level population. However, the
upper-level population given by (\ref{final}) is correct far from
the nodes of standing waves $g_1(x)$ and $g_2(y)$, where the
atomic population is almost entirely in the $|3\rangle$ state. On
the other hand, at the nodes of the standing waves expression
$g_1(x)=0$ or $g_2(y)=0$ is satisfied. Hence, the atomic
population at the nodes is concentrated either in the $|1\rangle$
state or in the $|2\rangle$ state and the upper-level population
at the nodes is zero.

Let us consider a running-wave case of the probe field, when
$g_3(x)=\Omega$. From (\ref{final}), (\ref{denom}) we selected
three types of distribution of the upper-level population and
called them spikes, craters and waves. These types can be seen in
Fig. \ref{upper} where the upper-level population is obtained by
numerically solving Eqs. (\ref{eq}) in the long-time limit.

One can see from (\ref{denom}) that the peaks of the spikes occur
in the $(x,y)$ positions given by
\begin{equation}
\sin^2 kx\approx 1,\quad\sin^2 ky\approx 1.
\end{equation}
These positions are
\begin{equation}\label{peaks}
(x,y)\approx(\:(2m+1)\lambda/4,\:(2n+1)\lambda/4\:)
\end{equation}
with $\lambda=2\pi/k$ and $m$, $n$ are integer. Thus, the spikes
give information how close the atom passes by the points
(\ref{peaks}) and in case of narrow spikes lead to localization of
the atom at these points. The craters in their turn allow one to
localize the atom at a distance from the positions (\ref{peaks}).
The positions of peaks of the craters follow from (\ref{denom})
and are derived from expression
\begin{equation}\label{peaks1}
\frac{G_1^2}{\Delta_{13}}\sin^2 kx+\frac{G_2^2}{\Delta_{23}}\sin^2
ky +\Delta_3\approx 0.
\end{equation}

In case of waves the positions of peaks are given by
(\ref{peaks1}) as well. In this case, by adjusting parameters of
the atom-field coupling, we can arrive at the condition
\begin{equation}
\left|G_2^2/\Delta_{23}\right|\ll\left|G_1^2/\Delta_{13}\right|.
\end{equation}
From (\ref{peaks1}) we then obtain that the expression
\begin{equation}\label{xfix}
\sin^2 kx\approx\Delta_3(\Delta_3-\Delta_1)/G_1^2
\end{equation}
is valid and the atom can be localized along the $x$ direction at
the positions (\ref{xfix}). Also in case of waves we can realize
the following condition
\begin{equation}
\left|G_1^2/\Delta_{13}\right|\ll\left|G_2^2/\Delta_{23}\right|,
\end{equation}
which leads to localization of the atom at the $y$ positions given
by
\begin{equation}\label{yfix}
\sin^2 ky\approx\Delta_3(\Delta_3-\Delta_2)/G_2^2.
\end{equation}

The degree of localization in case of spikes, craters or waves is
not limited by the optical wavelength and the width of peaks can
be essentially smaller than optical wavelength. A high-precision
localization is attained by increasing the Rabi frequencies $G_1$
and $G_2$, while the height of the peaks is defined by the
intensity $\Omega^2$ of the probe field.

\begin{figure}
\includegraphics[width=7.5cm]{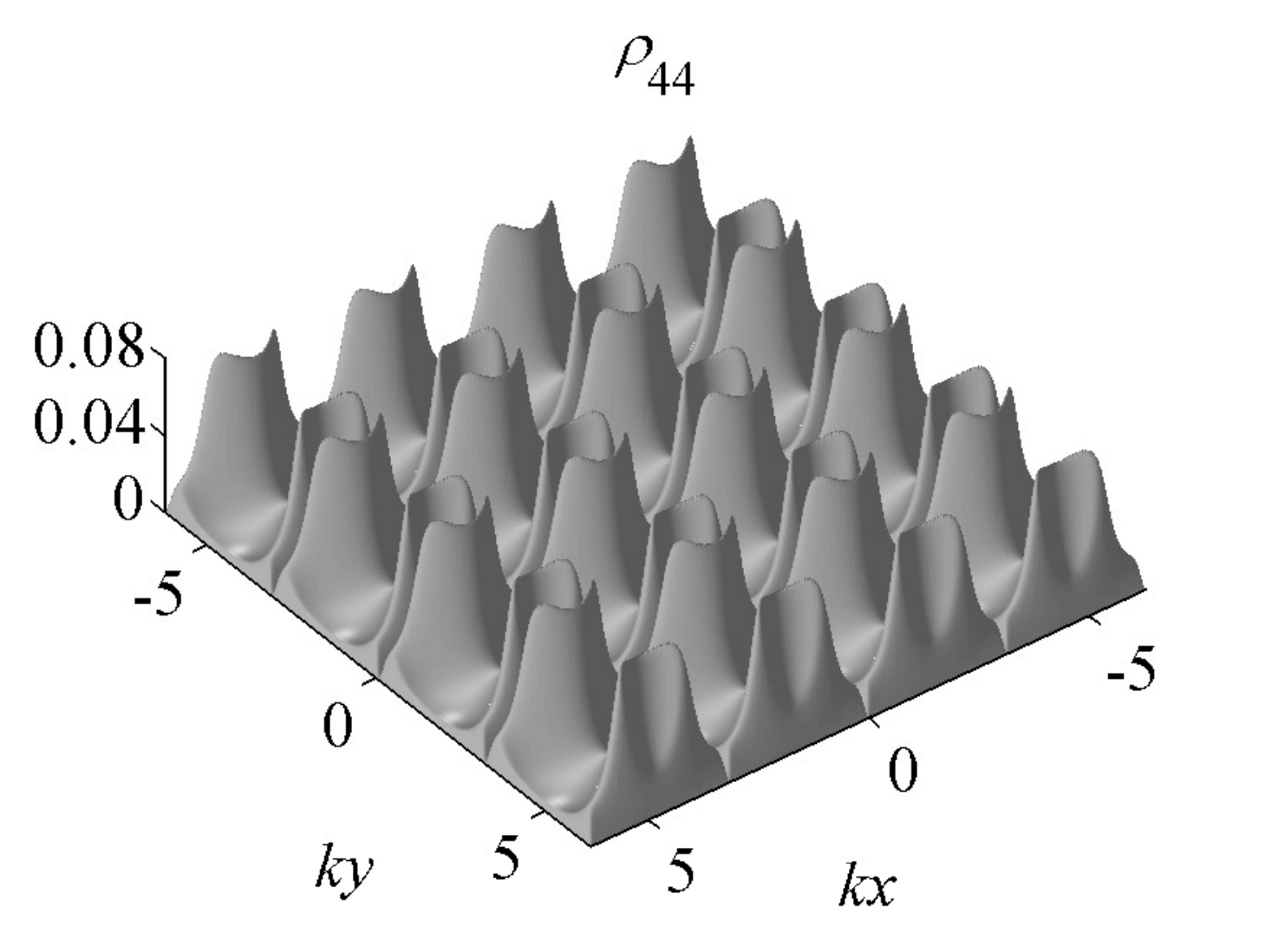}\\(a)\\
\includegraphics[width=7.5cm]{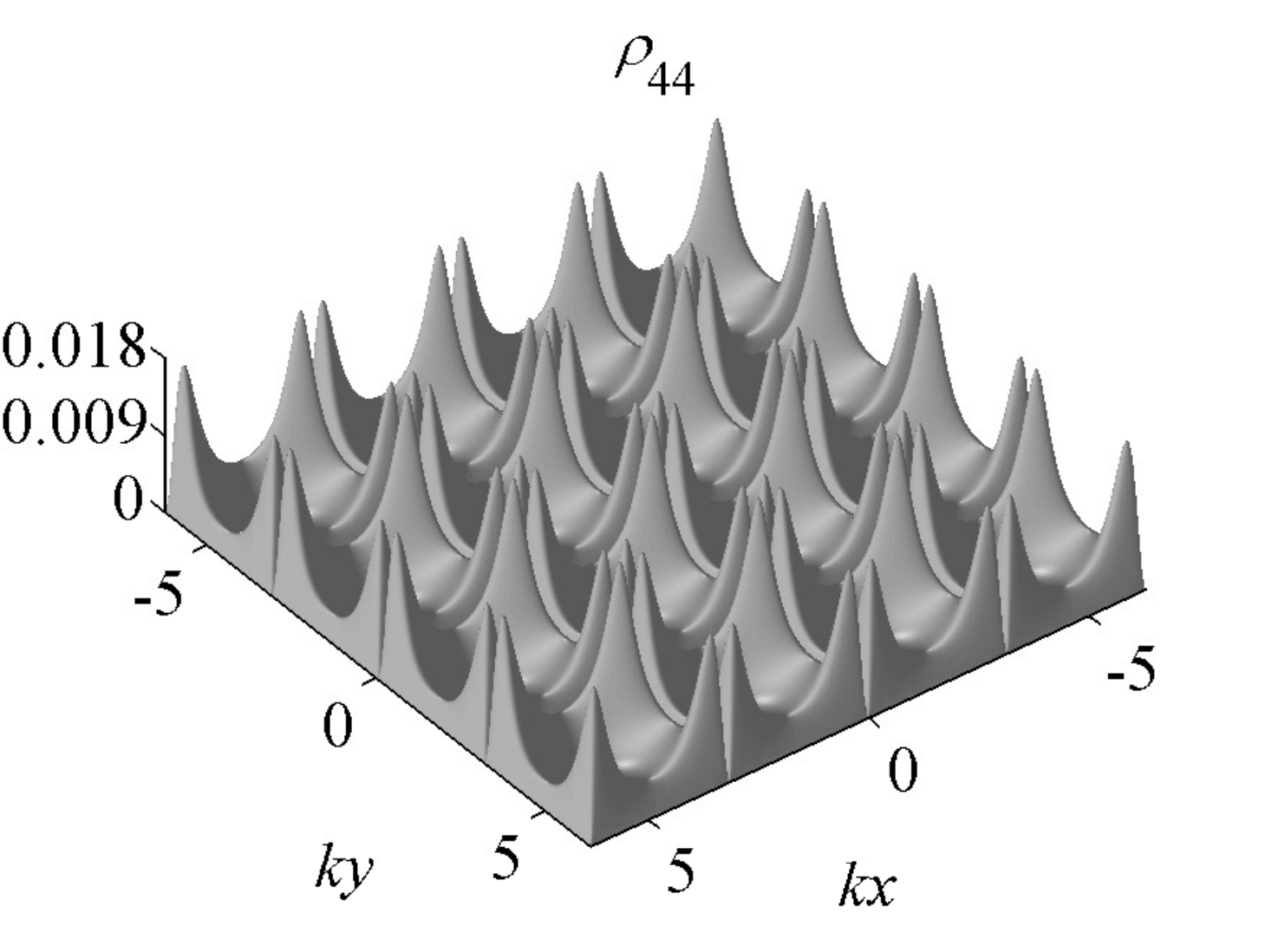}\\(b) \caption{\label{upper1}
Transferring the upper-level population by a spatial phase shift
of $\pi/2$ along the $x$ or $y$ direction: (a) craters shifted
along the $y$ direction to the nodes of standing wave $g_2(y)$;
(b) spikes shifted to crossings of the nodal lines. The laser
frequency detunings are (a) $\Delta_1=-3\gamma$,
$\Delta_2=5\gamma$, $\Delta_3=3\gamma$, and (b)
$\Delta_1=8\gamma$, $\Delta_2=4\gamma$, $\Delta_3=2\gamma$. The
Rabi frequencies of the standing waves are $G_1=6\gamma$,
$G_2=4\gamma$, and the Rabi frequency of the probe running wave is
$\Omega=0.3\gamma$. All other parameters are the same as in Fig.
\ref{upper}.}
\end{figure}

Let us show that the distribution of the upper-state population
can be transferred by a spatial phase shift of $\pi/2$ in the $x$
or $y$ direction. In a running-wave case of the probe field
spatial dependence of the upper-level population is defined by the
denominator (\ref{denom}). The following transformations
\begin{equation}
\begin{aligned}
\Delta'_1=&2\Delta_3-\Delta_1+G_1^2/\Delta_{13},\\
\Delta'_2=&\Delta_2+G_1^2/\Delta_{13},\\
\Delta'_3=&\Delta_3+G_1^2/\Delta_{13}
\end{aligned}
\end{equation}
convert this denominator to the form
\begin{gather}
Z'=\left(\frac{G_1^2}{\Delta'_{13}}\sin^2
kx+\frac{G_2^2}{\Delta'_{23}}\sin^2 ky
+\Delta'_3\right)^2+\Gamma_{34}^2\notag\\
=\left(\frac{G_1^2}{\Delta_{13}}\sin^2\left(kx\pm\frac{\pi}{2}\right)+
\frac{G_2^2}{\Delta_{23}}\sin^2
ky+\Delta_3\right)^2+\Gamma_{34}^2,\label{along_x}
\end{gather}
while the transformations
\begin{equation}
\begin{aligned}
\Delta''_1=&\Delta_1+G_2^2/\Delta_{23},\\
\Delta''_2=&2\Delta_3-\Delta_2+G_2^2/\Delta_{23},\\
\Delta''_3=&\Delta_3+G_2^2/\Delta_{23}
\end{aligned}
\end{equation}
change the denominator to the next form:
\begin{equation}\label{along_y}
Z''=\left(\frac{G_1^2}{\Delta_{13}}\sin^2 kx+
\frac{G_2^2}{\Delta_{23}}\sin^2\left(ky\pm\frac{\pi}{2}\right)+
\Delta_3\right)^2+\Gamma_{34}^2.
\end{equation}
One can see that the expressions (\ref{along_x}), (\ref{along_y})
describe the upper-level population (\ref{final}) shifted by a
spatial phase shift of $\pi/2$ along either the $x$ or $y$ axis.

Figure \ref{upper1} depicts the upper-level population obtained
numerically from Eqs. (\ref{eq}) that illustrates the result of
employing the transformations (\ref{along_x}), (\ref{along_y}).
Figure \ref{upper1}(a) shows craters after shifting along the $y$
direction to the nodes of standing wave $g_2(y)$ and Fig.
\ref{upper1}(b) shows spikes after shifting to crossings of the
nodal lines. These structures are located near the nodes of the
standing waves and take a new form due to crossing by the nodal
lines.

In a standing-wave case of the probe field, when
$g_3(x)=\Omega\sin(kx+\psi)$, the upper-state population
essentially decreases near the nodes of the probe field. On the
other hand, by adjusting the spatial phase shift $\psi$, most of
the upper-level population can be concentrated far from the nodes
of the probe field. Then, distribution of the population takes the
form similar to the running-wave case of the probe field and the
spatial structures look like those in Figs. \ref{upper},
\ref{upper1}.

\section{2D atom localization via a ground-level population}
\label{s4}

\begin{figure}
\includegraphics[width=7.5cm]{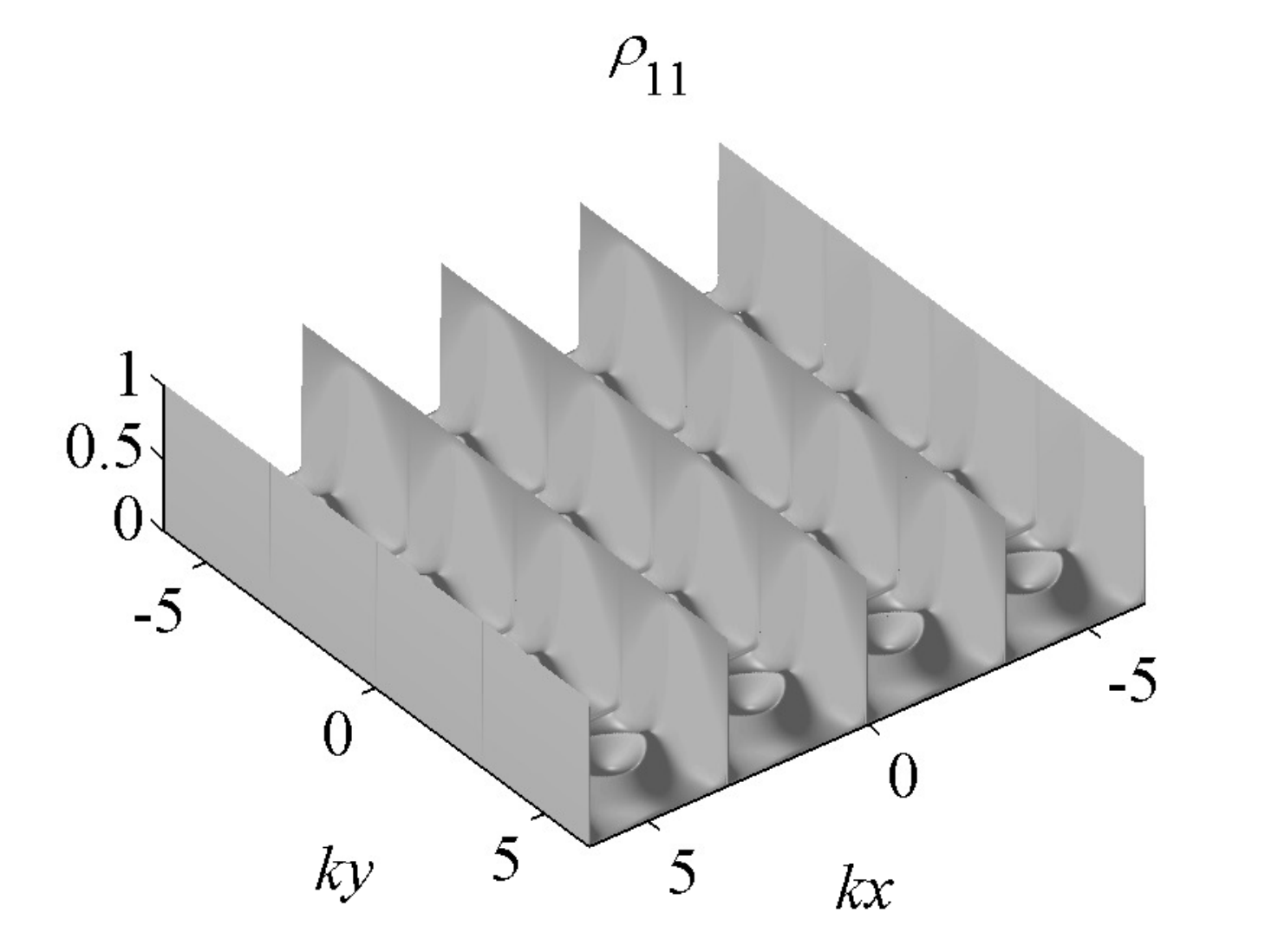}\\(a)\\
\includegraphics[width=7.5cm]{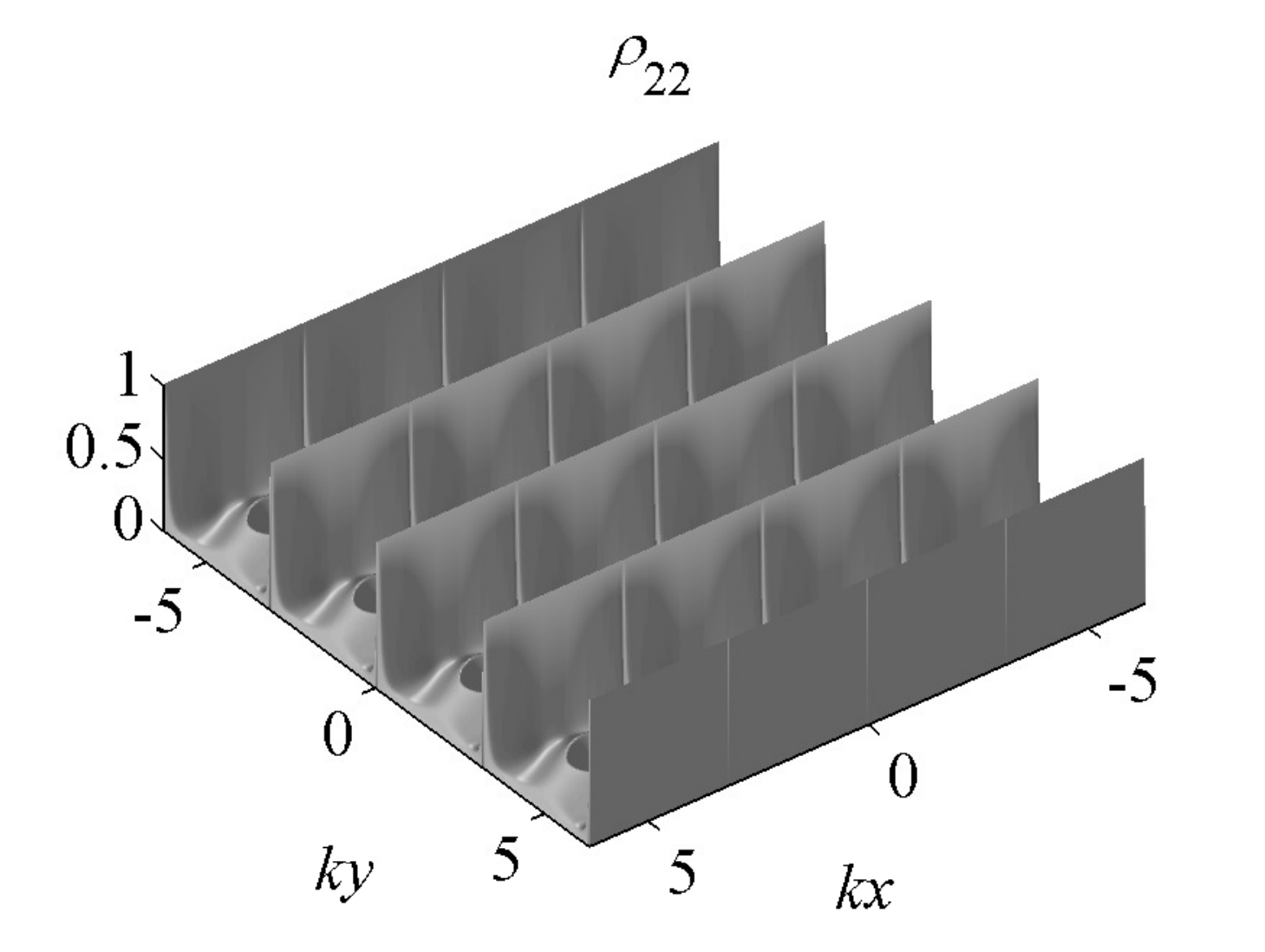}\\(b)\\
\includegraphics[width=7.5cm]{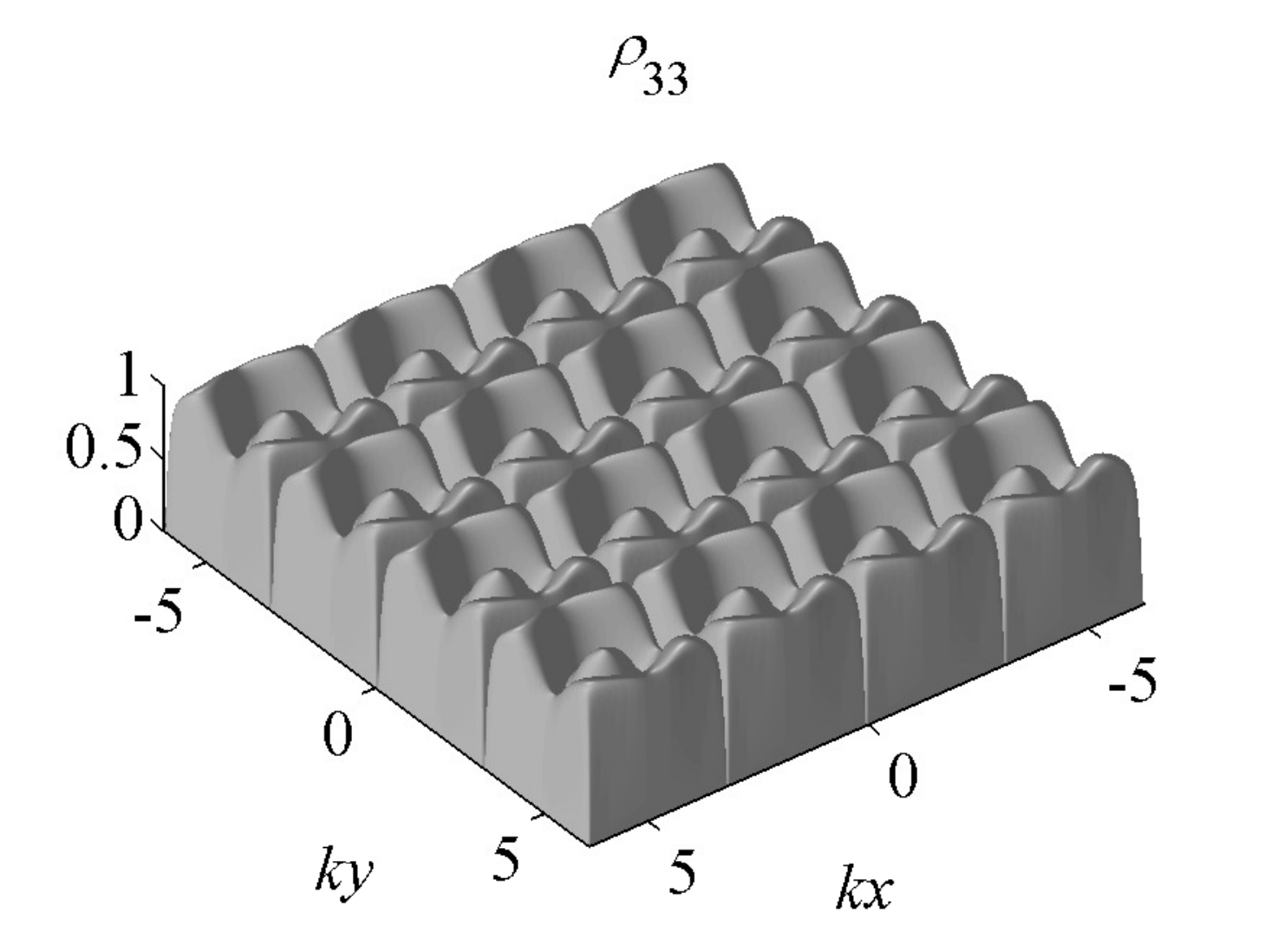}\\(c) \caption{\label{ground}
The ground-state populations as functions of $(kx,ky)$. (a)
Population in state $|1\rangle$ has narrow peaks along the nodal
lines of standing wave $g_1(x)$; (b) population in state
$|2\rangle$ has narrow peaks along the nodal lines of standing
wave $g_2(y)$; (c) population in state $|3\rangle$ has narrow
peaks along the nodal lines of both standing waves. The laser
frequency detunings are $\Delta_1=5\gamma$, $\Delta_2=9\gamma$,
$\Delta_3=11\gamma$; the Rabi frequencies of the standing waves
are $G_1=6\gamma$, $G_2=4\gamma$, and the Rabi frequency of the
probe running wave is $\Omega=0.3\gamma$. All other parameters are
the same as in Fig. \ref{upper}.}
\end{figure}

\begin{figure}
\includegraphics[width=7.5cm]{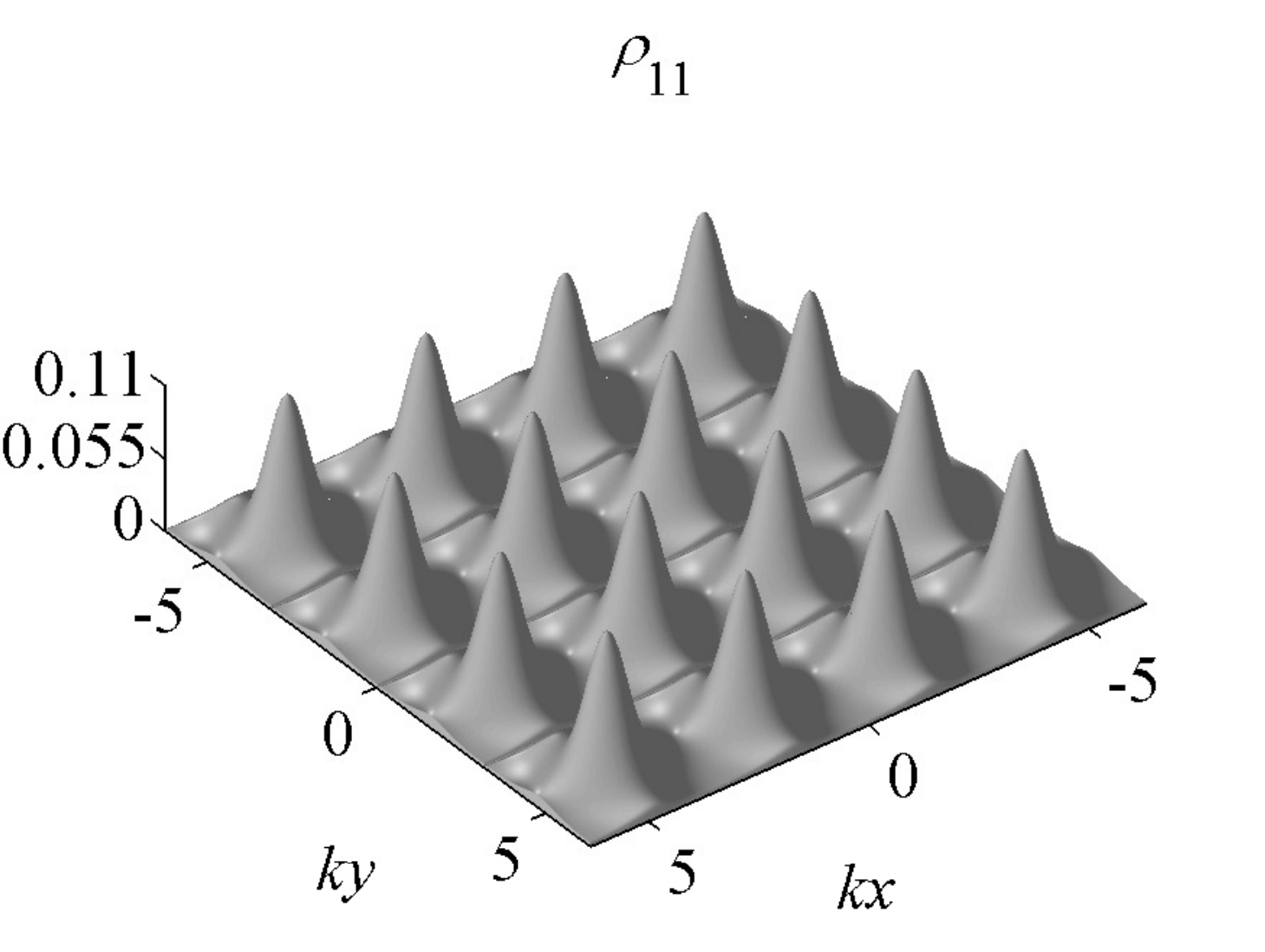}\\(a)\\
\includegraphics[width=7.5cm]{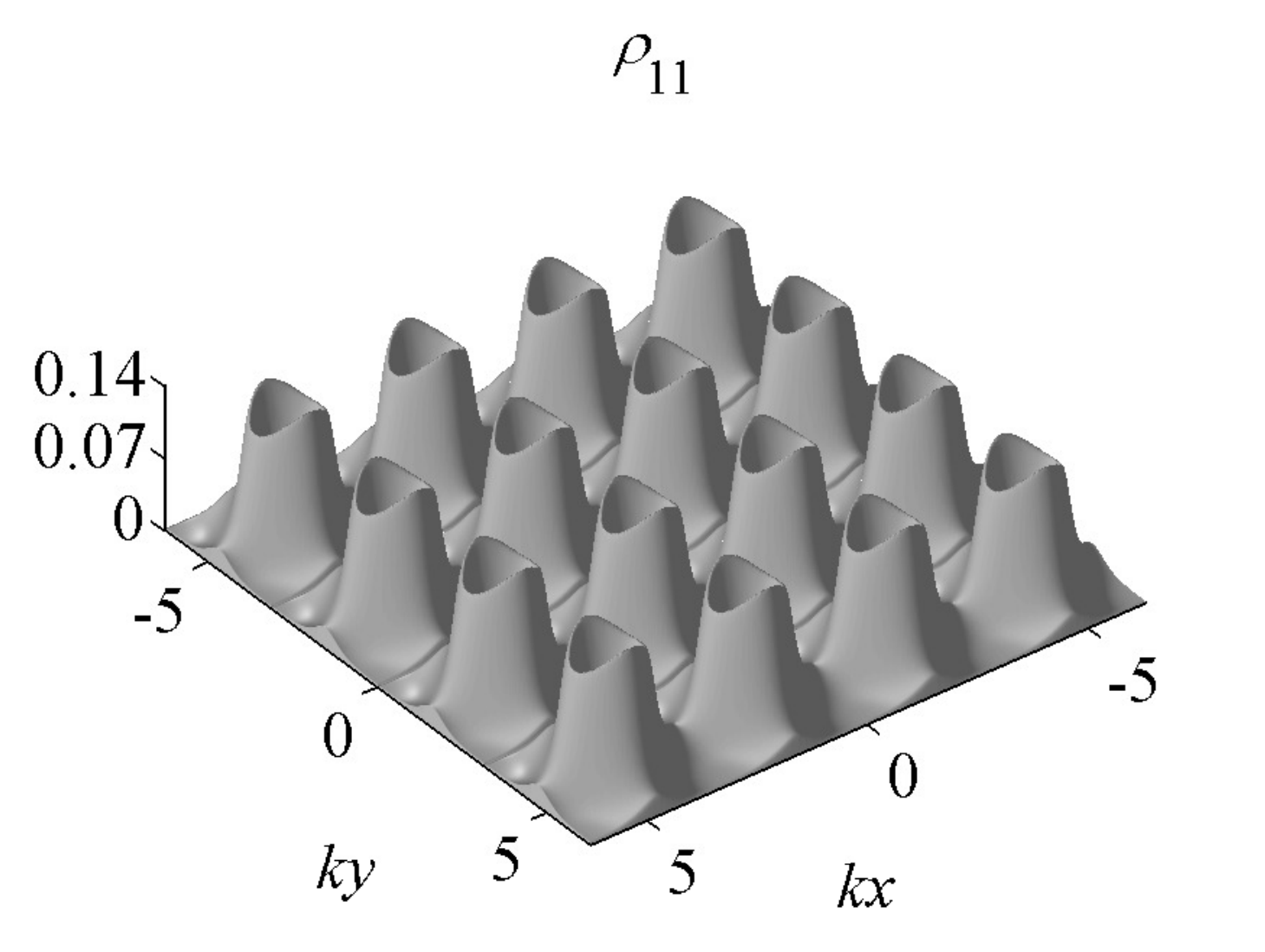}\\(b)\\
\includegraphics[width=7.5cm]{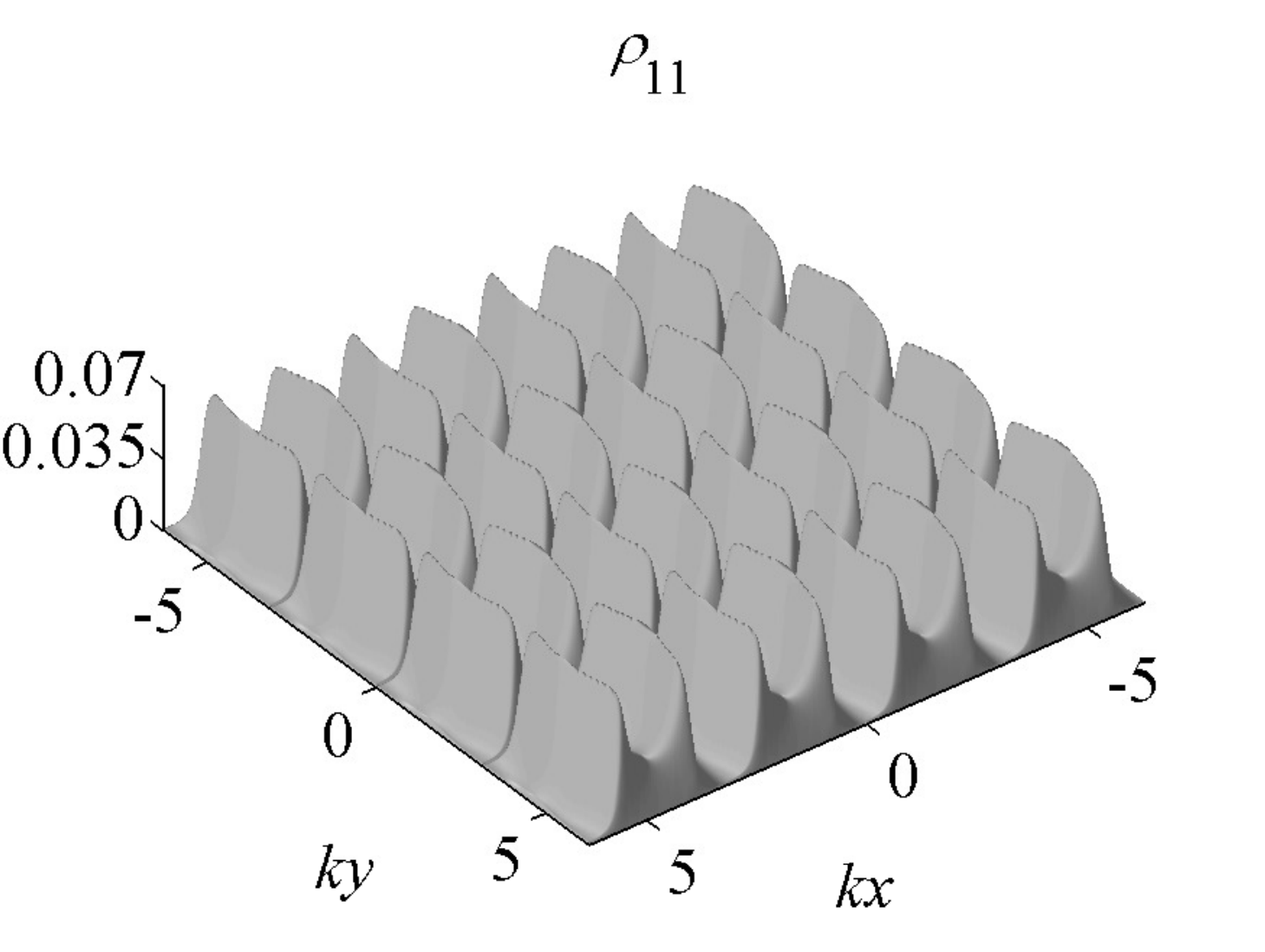}\\(c) \caption{\label{state1}
The population in ground state $|3\rangle$ as a function of
$(kx,ky)$ in limit case $\psi=0$. The spatial distributions of the
population are similar to those of population in the upper-state
and are (a) spikes, (b) craters and (c) waves depending on the
laser frequency detunings, (a) $\Delta_1=10\gamma$,
$\Delta_2=14\gamma$, $\Delta_3=16\gamma$; (b) $\Delta_1=5\gamma$,
$\Delta_2=9\gamma$, $\Delta_3=11\gamma$; and (c)
$\Delta_1=\gamma$, $\Delta_2=-3\gamma$, $\Delta_3=5\gamma$. The
Rabi frequencies of the standing waves are $G_1=6\gamma$,
$G_2=4\gamma$, and the Rabi frequency of the probe standing wave
is $\Omega=0.3\gamma$. All other parameters are the same as in
Fig. \ref{upper}.}
\end{figure}

\begin{figure}
\includegraphics[width=7.5cm]{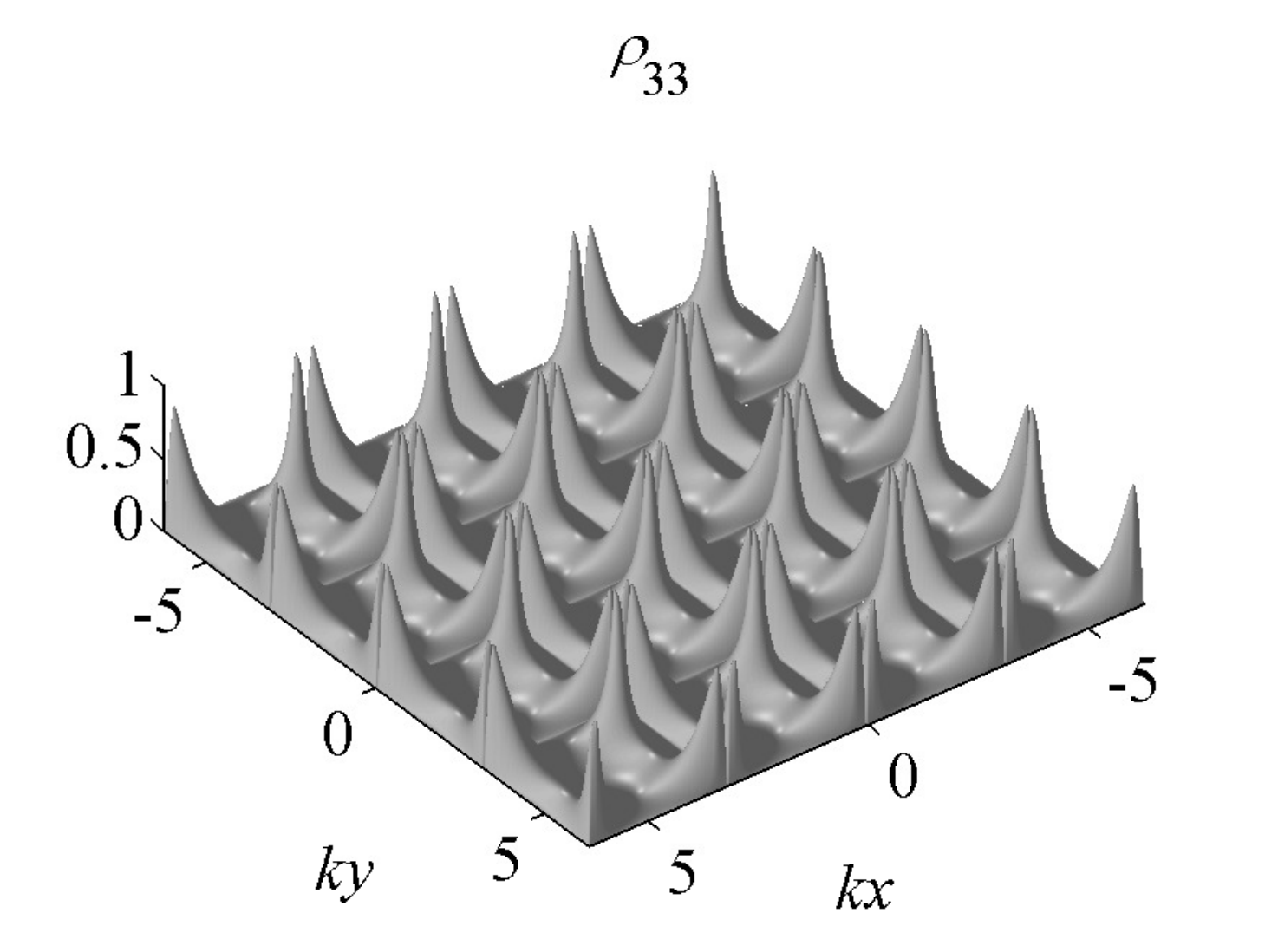}\caption{\label{state3}
Spatial distribution of population in the $|3\rangle$ state under
condition $\Delta_1\approx\Delta_2$. Peaks of the population are
located near crossings of the nodal lines, width of the peaks is
decreased by drawing near to CPT. The laser frequency detunings
are $\Delta_1=10\gamma$, $\Delta_2=10.1\gamma$,
$\Delta_3=15\gamma$; the Rabi frequencies of the standing waves
are $G_1=G_2=5\gamma$, and the Rabi frequency of the probe running
wave is $\Omega=0.3\gamma$. All other parameters are the same as
in Fig. \ref{upper}.}
\end{figure}

Localization of an atom can be performed by means of measuring the
population whether in the upper state or in a ground state. These
populations are defined by the atom-field interaction and they are
rather different for each internal state. Therefore, we can choose
not only parameters of the external fields, but we can also choose
an internal level which is more suitable for measurement. In this
section we analyze populations obtained numerically from Eqs.
(\ref{eq}) in the long-time limit and demonstrate a high-precision
localization due to measuring a ground-state population.

Distributions of the ground-state populations in a running-wave
case of the probe field are shown in Fig. \ref{ground}. Peaks of
population in the $|1\rangle$ state occur at the nodes of standing
wave $g_1(x)$ (see Fig. \ref{ground}(a)), and peaks of population
in state $|2\rangle$ occur at the nodes of standing wave $g_2(y)$
(see Fig. \ref{ground}(b)). The atomic population at these peaks
is concentrated in one ground state, so the atom can be localized
much more easily than for the upper state. Population in the
$|3\rangle$ state shown in Fig. \ref{ground}(c) is avoided at the
nodes of both standing waves, $g_1(x)$ and $g_2(y)$. Hence, in
this way we have another type of localization when the atom due to
absence of population can be localized at the nodes of both
standing waves. Peaks of population along the $x$ axis have the
width defined by the ratio of $\Omega/G_1$, while peaks of
population along the $y$ axis have the width defined by the ratio
of $\Omega/G_2$. The high-precision localization is obtained by
means of decreasing the ratios of $\Omega/G_1$ and $\Omega/G_2$.

Let us consider a standing-wave case of the probe field. The width
of the peaks along the $x$ direction can be decreased not only by
means of the ratio of $\Omega/G_1$. In addition, the small width
of the peaks is obtained when the nodes of the probe field are
approximate to the nodes of standing wave $g_1(x)$, i.e. spatial
phase shift $\psi\approx 0$. In the case of $\psi=0$ the peaks of
the population in the $|1\rangle$ state are avoided, which leads
to spatial structures of the population shown in Fig.
\ref{state1}. Distributions of the population take the forms
similar to those of population in the upper state and can
represent spikes, craters and waves. The heights of these
structures are defined by the intensity $\Omega^2$ of the probe
field.

Localization conditions discussed earlier correspond to large
differences between the laser frequency detunings $\Delta_1$,
$\Delta_2$ and $\Delta_3$. On the other hand, it is a well-known
fact that in a tripod-like atom coherent population trapping (CPT)
takes place when two of the laser detunings are equal. The 1D
localization of an atom based on CPT has been demonstrated by
Agarwal \textit{et al.} \cite{Ag} for a three-level $\Lambda$-type
atom due to measurement of a ground-state population. Here we
propose a way to carry out the 2D atom localization in a
tripod-type atom under a condition close to CPT.

Let us consider the case of the running-wave probe field and
$\Delta_1\approx\Delta_2$. While the atomic population under usual
conditions is almost entirely in the $|3\rangle$ state, the atomic
population in case of $\Delta_1=\Delta_2$ is trapped in the
$|1\rangle$ and $|2\rangle$ states. So, when
$\Delta_1\approx\Delta_2$ the behavior of the atomic populations
rather differs from other cases and is of special interest. Note
that the strong standing waves push out the atomic population from
states $|1\rangle$ and $|2\rangle$ and hinder CPT, so intensities
of the standing waves should be moderate. The distribution of
population in the $|3\rangle$ state is shown in Fig. \ref{state3}.
Peaks of the population are located near crossings of the nodal
lines. Width of the peaks is decreased by drawing near to CPT, and
the high-precision localization is achieved for the small value of
$|\Delta_{12}|$.

\section{Conclusion}
\label{s5}

In conclusion, we have presented a scheme for 2D subwavelength
localization of a four-level tripod-like atom in laser fields. 2D
spatial periodic structures of the upper-state population such as
spikes, craters and waves are selected. The same structures are
observed for population in ground state $|1\rangle$ in the case of
$\psi=0$. Also, the atom observed in a ground state can be
localized at the nodes of one of the standing waves. As a result,
we have demonstrated methods having important practical
applications in high-resolution atom lithography that are of
particular interest because of the recent experiments for
semiconductor elements \cite{Lee,Chen,Ind}.

\section*{ACKNOWLEDGMENTS}

This research was supported by the Ministry of Education and
Science of the Russian Federation under Grant RNP 2.1.1/2166, and
by the Russian Foundation for basic research, Grant RFBR
09-02-00223a. It was also supported by the Magnus Ehrnrooth
Foundation, and by the Academy of Finland, Grant 115682.

\end{document}